*Alexander Mikhailichenko, July 1 2015*

# FEASIBILITY OF PULSED INFLECTOR FOR g-2 RING (1)

Abstract. We revise a possibility of usage iron-free fast pulsed magnet as inflector for g-2 ring, coming to conclusion that this is a feasible option.

## INTRODUCTION

Variant with pulsed inflector, which is declined in E-989 TDR, states, that the inflector magnet "*should be a static device to prevent time-varying magnetic fields correlated with injection*, which could affect $\oint \vec{B} d\vec{l}$ seen by the stored muons and produce an "early to late" systematic effect". It should be protected against field leak onto orbit to avoid distortion of $\oint \vec{B} d\vec{l}$ at sub-ppm level [1].

However one pulsed magnet, namely the *kicker* in g-2 ring, is acting on the beam orbit correlated with injection. This means that restrictions put in grounds of inflector design, beginning from E-821, should be revised or specified more carefully at least.

The general function of inflector magnet is in cancellation of main magnetic field (~1.5 *T*) on the trajectory of particle on its way from outside of the ring to the location of the edge of allowable aperture of main ring, which is ~77*mm* off the equilibrium radius of orbit which is 280 inches big.

## THE CONCEPT

Basic concept of the pulsed inflector, which we are suggesting here, is represented in Fig.1. In general, the magnet is an asymmetric strip-line feed directly by discharge of capacitor through thyristor(s) directly to the inflector magnet or through a transformer. To prevent leakage of field onto orbit, the strip-line magnet enclosed by a thick-wall copper container, so the skin-effect serves for the purposes of attenuation. The wall thickness of this container should be big enough for attenuation of field by skin-effect. For example, if duration of the current pulse in a magnet is, say 20 *μs*, then attenuation of the field outside of (copper) container having thickness 1*cm* comes to $e^{-31}$, where it was taken into account that for 10*ms* half-pulse duration (50*Hz*) the skin-layer is $d_s \approx \sqrt{2\rho ms / \omega \mu_r \mu_0} \approx 1$ *cm* (ρ stands for resistivity of conductor) and for 20 *μs* it comes to $d_s \approx 1 cm \sqrt{20 m s / 20 ms} \approx 1/31.6$ and the attenuation $e^{-x/d_s} \approx e^{-31.6}$ ~1.9x10$^{-14}$. Vertical dimension of container is limited by the (vertical) dimension of the vacuum chamber, which fits between the poles of main magnet of g-2 ring, ~16.5 *cm* (6.5") and its walls.

Driving conductor shape in a transverse direction is shown as a rectangular in Figs.1,2. It is clear, that the profile of conductor could be modified at the top and bottom for obtaining higher homogeneity of field in aperture of inflector. One can see that attenuation is so big, that even longer pulses become allowable, taking into account a square-root dependence of skin-depth on the pulse duration. Say, even for 100 $\mu s$ pulse duration, skin-layer will be $d_s \approx \sqrt{5}/31.6 \approx 0.07 [cm]$ and attenuation arrives at $e^{-x/d_s} \approx 7.3 \times 10^{-7}$.

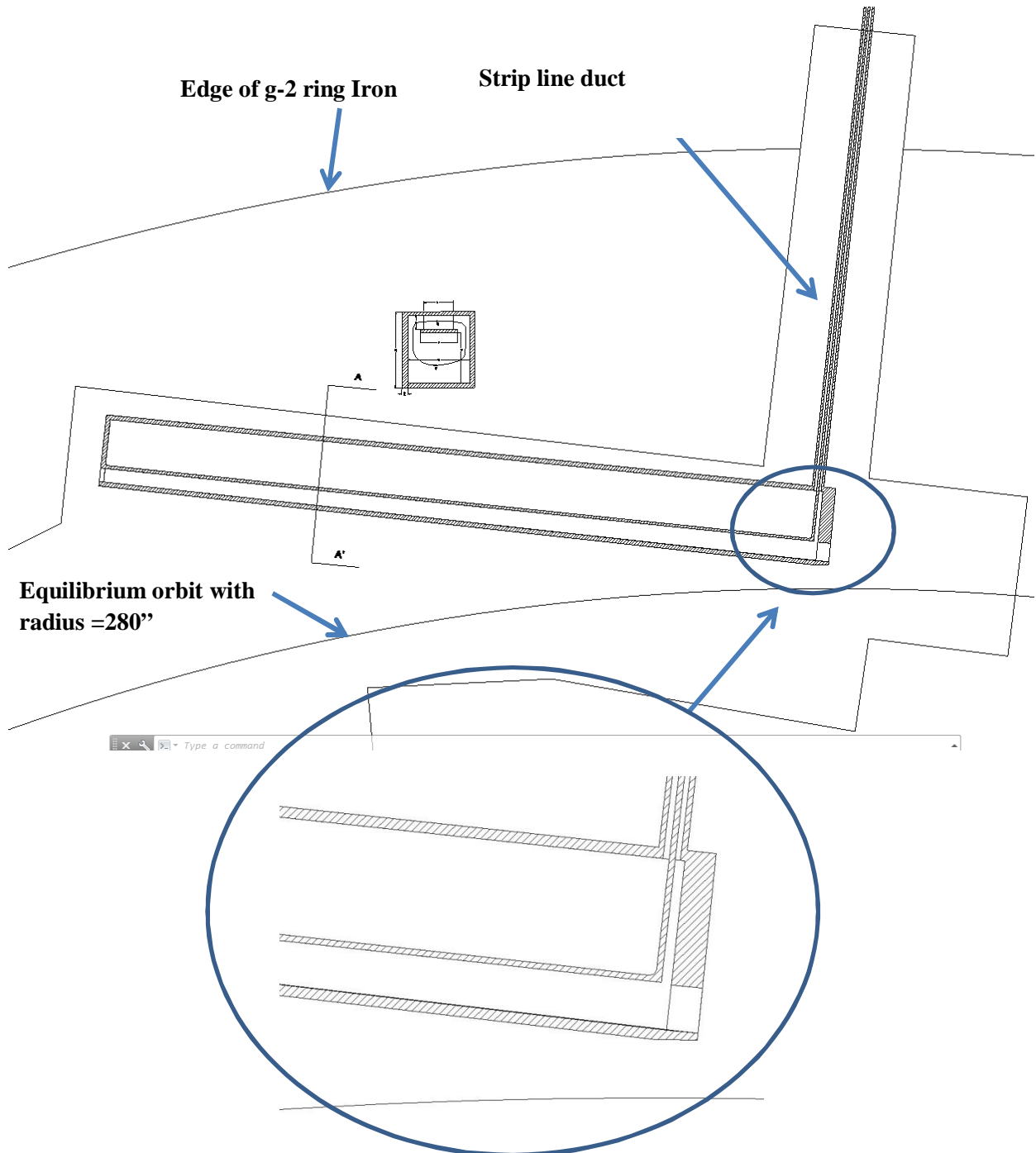

**Figure1.** The concept. Thick end-wall introduced for prolongation of orifice with aperture shape for exponential attenuation of magnetic field leakage into main vacuum chamber through the aperture.

Taking into account that the thickness of wall of Al chamber of g-2 ring is 1.33 *cm*, one can allow the vertical dimension of container $H+2 \times del = 13.84$ *cm*, where *del* stands for the thickness of walls of container, so *H* =11.84 *cm*, if the thickness of container is *del*=1*cm, see* Fig.2. Here we took into account that the total height of Al chamber is 6.5" =13.85*cm*. In its turn as the vertical size of central conductor chosen to be *h*~5*cm*, then the vertical distance between the central conductor and the inner wall of container (*H-h*)/2, comes to (11.84-5)/2=3.42 *cm*. Dimension *h*=5 *cm* defined by vertical aperture of inflector magnet, which was suggested to be *a*~4*cm* and *d*=2*cm*. The exact value of *h* will be identified by numerical calculation by requirements of homogeneity of field in aperture of inflector magnet.

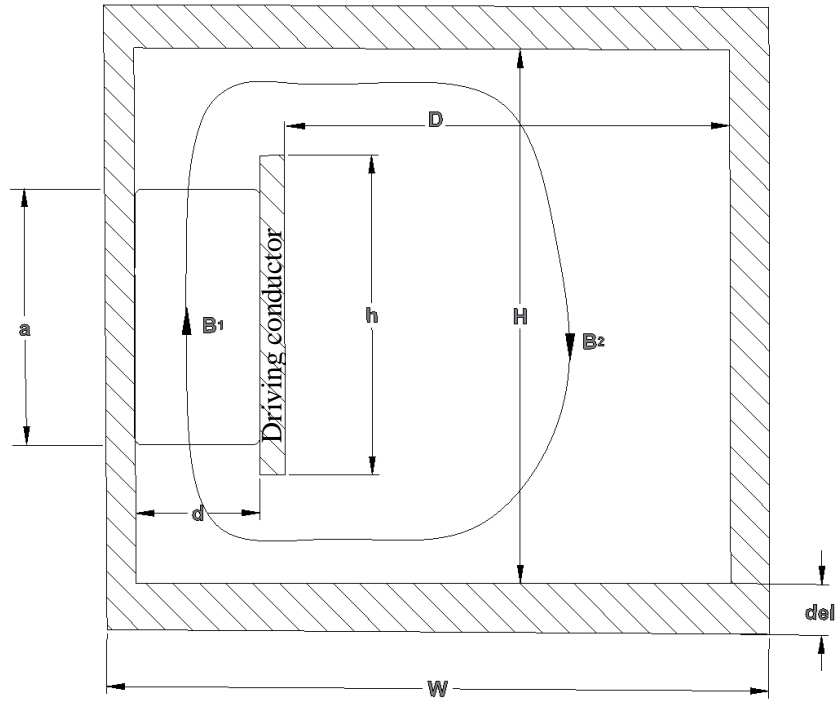

**Figure 2**. Dimensions of strip line magnet used for evaluation of parameters of PS.

We take care of this distance as we are trying to minimize reduction of field in the aperture of inflector as the flux of magnetic field is conserved in transverse cross-section. So the places with transversely squeezed flux deliver the biggest input in in field integral. That is why the vertical dimension of container above the driving conductor in Fig. 2 (having dimension *h*) was chosen as big as possible, so the field input in the integral $\oint \vec{B} d\vec{l}$ around the driving conductor arisen at the distance *h* mostly.

## ENERGETICS OF INFLECTOR MAGNET

Current defined by the desirable field could be calculated as the following. As we intend to cancel the main field of g-2 ring magnet, then inflector should generate the field~1.5*T* of opposite direction, which is the level of field present on the beam orbit. Now it is possible to evaluate the current required by $\oint \vec{B} d\vec{l} = \mu_0 \times I$, where integration is going along the line shown in Fig.2 in transverse direction. One other condition which should be taken into account associated with conservation of flux of magnetic field in transverse direction

$$B_1 \times d \approx B_2 \times D \tag{1}$$

Where $B_1$ is the field in aperture of inflector and $B_2$ is the field at the other side of strip-line, Fig.2. As we've chosen $D \sim 5d$, then $B_2 \sim 1/5 B_1$ and identity $\oint \vec{B} d\vec{l} = \mu_0 \times I$ comes to .

$$\oint \vec{B} d\vec{l} \approx B_1 \times h + B_2 \times h = \mu_0 I \Rightarrow (B_1 + \tfrac{1}{5} B_1) \times h = \mu_0 I, \tag{2}$$

so

$$B_1 = \tfrac{5}{6} \frac{\mu_0 I}{h}. \tag{3}$$

As $B_1 \sim 1.5 T$, then taking into account that $h = 5 cm$ we obtain the current value required

$$I \approx \tfrac{6}{5} \frac{B_1 \times h}{\mu_0} \approx \tfrac{6}{5} \frac{1.5 \times 5 \times 10^{-2}}{4\pi \times 10^{-7}} \approx 0.75 \times 10^5 = 75 [kA]. \tag{4}$$

In our further estimation of power required we will use $I = 100 kA$ as a reference current value, however.

For calculation of inductance of magnet $L_{inf}$ we can recall an identity

$$L_{inf} I^2 \approx \oiiint \frac{B_1^2 \times dV}{\mu_0} \approx \frac{B_1^2 \times h \times d \times l}{\mu_0} \approx 1.4 \mu_0 \frac{I^2 \times d \times l}{h}, \tag{5}$$

where $l \sim 1.5\ m$ stands for the length of inflector magnet. So inductance comes to

$$L_{inf} \approx 1.4 \mu_0 \frac{d \times l}{h} = 1.4 \times 4\pi 10^{-7} \frac{2 \times 10^{-2} \times 1.5}{5 \times 10^{-2}} \approx 1 \times 10^{-6} [Henry] \tag{6}$$

If we suggest, that the pulse duty has a shape of half-sine wave with $\tau \approx 20\ ms$ from zero to zero, then the impedance of magnet comes to

$$Z \approx \frac{2\pi}{2\tau} L \approx \frac{2\pi}{40 \times 10^{-6}} 1.7 \times 10^{-7} \approx 2.7 \times 10^{-2} [\Omega], \tag{7}$$

So the voltage applied to the strip-line for 100 $kA$ current will be

$$V = IZ \approx 10^5 \times 2.7 \times 10^{-2} [\Omega] = 2.7 [kV] \tag{8}$$

Inductance of the feeding strip-line duct could be calculated as the following. First, as the voltage is not high, so insulation such as Kapton tape will be adequate. Let the height of central conductor be $\Delta = 10 cm$ and the thickness of insulation let be $\delta = 1 mm$. Magnetic field between central conductor and surrounding wall comes to $B_d = \frac{\mu_0 I}{2D}$. Then the energy accumulated in magnetic field of duct comes

$$LI^2 \approx \frac{\oiiint B_d^2 \times dV}{\mu_0} \approx 2\frac{B_d^2 \times d \times L \times l_d}{\mu_0} \approx 2\mu_0 \frac{I^2 \times d \times l_d}{4D} , \qquad (9)$$

where $l_d$ is the length of stripline duct, $l_d \sim 2m$, and the factor 2 reflects the fact that for symmetrical stripline the energy stored at both sides of central conductor. So the inductance of duct line comes to

$$L \approx 2\mu_0 \frac{d \times l_d}{4D} = 2 \times 4\pi \times 10^{-7} \frac{10^{-3} \times 2}{4 \times 0.1} = 4\pi \times 10^{-9} \approx 12.5 [nH] , \qquad (10)$$

which is negligible if compared with the inductance of main magnet (5). So the secondary voltage majored by (8).

Energy accumulated in magnetic field is

$$W \approx \tfrac{1}{2} LI^2 \approx \tfrac{1}{2} 10^{-6} 10^{10} = 5 kJ . \qquad (11)$$

As the average repetition rate in g-2 complex is 12 *Hz*, the average power comes to 60 *kJ/s* or 60*kW*. With recuperation, this amount will be lower, see section *Power Supply*.

## FORCES

As the driving conductor (see Fig.2) immersed into $B_{ext}=1.5T$ field, it experiences substantial magneto-force, which could be calculated as

$$\vec{F} \approx \oint \vec{j} \times \vec{B}_{ext} dV \approx \vec{n} \times I \times B_{ext} \times L , \qquad (12)$$

where $\vec{n}$ is unit vector in normal to the surface of conductor direction. Substitute in (12) the current value $I=100kA$, the field value $B_{ext}=1.5T$ and for the length $l=1m$, one can obtain that $F \approx I \times B_{ext} \times L = 100[kA] \times 1.5[T] \times 1[m] = 150kN = 15T$. This force acts for $\tau \sim 20\mu s$ however with sin-like shape. Taking into consideration that the mass of $1m$-long piece of driving electrode is $m \sim 1kg$ the acceleration is $a = F/m \approx 1.5 \times 10^4 [m/s^2]$. The distance passed by conductor becomes equal to

$$S \approx \frac{at^2}{2} = \frac{1.5 \times 10^4 \times 20 \times 20 \times 10^{-12}}{2} = 3 \times 10^{-6}[m] \approx 3 \mu m . \qquad (13)$$

Such displacement one can expect if conductor is not surrounded by container. From the other hand as the direction of driving current is chosen so, that in the aperture of inflector the field vanishes, the magnetic pressure now acts to the conductor directed towards aperture of inflector, as at outer volume field remains $B_1=1.5T$ (or even slightly elevated, in accordance with (1)). So we can estimate the pressure as

$$P \approx \frac{B_1^2}{2\mu_0} = \frac{1.5 \times 1.5}{8\pi} 10^7 \approx 9 \times 10^5 [N/m^2] \qquad (14)$$

So the force acting on the conductor is

$$F = P \times h \times l \approx 9 \times 10^5 \times 5 \times 10^{-2} \times l \approx 5 \times 10^4 [N] \quad @ \; 5T \qquad (15)$$

We should conclude, that namely this value, which is ~3 times lower, than calculated with (12), should be taken as a final one. So the displacement comes to $S \approx 1mm$. It is interesting, that if the field inside main magnet is turned off, the magnetic pressure reverses its sign and acts in outside direction (out of inflector aperture). Although the value of displacement is small enough, the central conductor should be supported adequately, for prevention of movement in both directions. Non- magnetic materials could be used for this purpose, like G11 as radiation exposure by the muon beam is negligible here.

## POWER SUPPLY

*Direct feeding with high-current fast thyristor(s)* is a preferable option. There are many publications on practical utilization of thyristors for fast commutation of high currents [2]-[5]. The last publication, [5], describes a single thyristor able to commute up to 500 *kA*.

One good candidate for usage in our inflector is thyristor switch 5STH 20H4501 by ABB Switzerland Ltd, semiconductors. It holds DC voltage 2.8 *kV*, repetitive peak of-state voltage 4.5 *kV* and allows pulsed current 80 *kA* with maximal current rate of rise $dI/dt$=18 *kA/μs*, see Addendum. With adequate cooling the repetition rate could be 100*Hz*, as required for our purpose (more exactly the average repletion rate is 12 *Hz*, as the pulses are grouped in two trains by 8 pulses in each one separated by 10 *ms,* following with ~197*ms and* 1063*ms* intervals covering ~1.4 *sec* cycle).

For the value of capacity required we can evaluate that $T = \frac{1}{2\pi}\sqrt{LC}$, $T = 2 \times 20 ms$ so

$$C = \frac{1}{L}(2\pi \cdot T)^2 = \frac{1}{10^{-6}}(4\pi \cdot 20 \times 10^{-6})^2 \approx 250 \; [mF]. \qquad (16)$$

So that might be a single low inductance capacitor or few ones in parallel.

Electric scheme is represented in Fig.3. Here for charging capacitor we suggest a specially designed for these purposes PS 303 series from TDK-Lambda (Electronic Measurements Inc.) This PS allows controllable pulsed charge of capacitors with power up to 37.5 *kJ/s* (50*kW* of average power). These PS could work in parallel.

In principle single thyristor can carry 75 *kA* estimated in (4), one another thyristor could be attached in parallel, however. In last case some matching impedances should be connected in series with each branch for alignment of currents. As the current which is running in a circuit is practically within specification that will be not a problem.

Controllable recharge of capacitor (recuperation) is going through additional inductance $L_r$ in series with additional thyristor. This last one could be with much modest parameters as the inductance $L_r \gg L_{inf}$. The moment of re-charging could be controlled by this thyristor; this option might be useful.

The scheme is basically a LCR circuit which characteristic frequency $f$ is defined by the expression

$$f \approx \frac{1}{2\pi}\sqrt{\frac{1}{LC} - \frac{R^2}{4L^2}}. \tag{17}$$

So if the expression under the square root is negative, the discharge becomes aperiodic. Minimal value of $R$ could be found from identity

$$\frac{L}{C} = \frac{R^2}{4}, \tag{18}$$

so

$$R = 2\sqrt{\frac{L}{C}} \approx 2\sqrt{\frac{10^{-6}}{250 \times 10^{-6}}} = 0.126\,[\Omega]. \tag{19}$$

The resistance is pretty small which is explained by low characteristic wave resistance

$$R_w = \sqrt{\frac{L}{C}} \approx \sqrt{\frac{10^{-6}}{250 \times 10^{-6}}} = 0.063\,[\Omega]. \tag{20}$$

So the quality factor for these parameters becomes

$$Q = \frac{1}{R}\sqrt{\frac{L}{C}} \approx \frac{1}{0.063 \times 15.8} \gg 1, \tag{21}$$

this means that the discharge is aperiodic. For the scheme with thyristor a peculiarity is that thyristor does not conduct the current in opposite direction (except special ones).

This means, that inductance value does not playing decisive role in this case, and discharge is going like a simple RC one with the time constant $\tau = RC = 0.126 \times 250 = 31\,[\mu s]$.

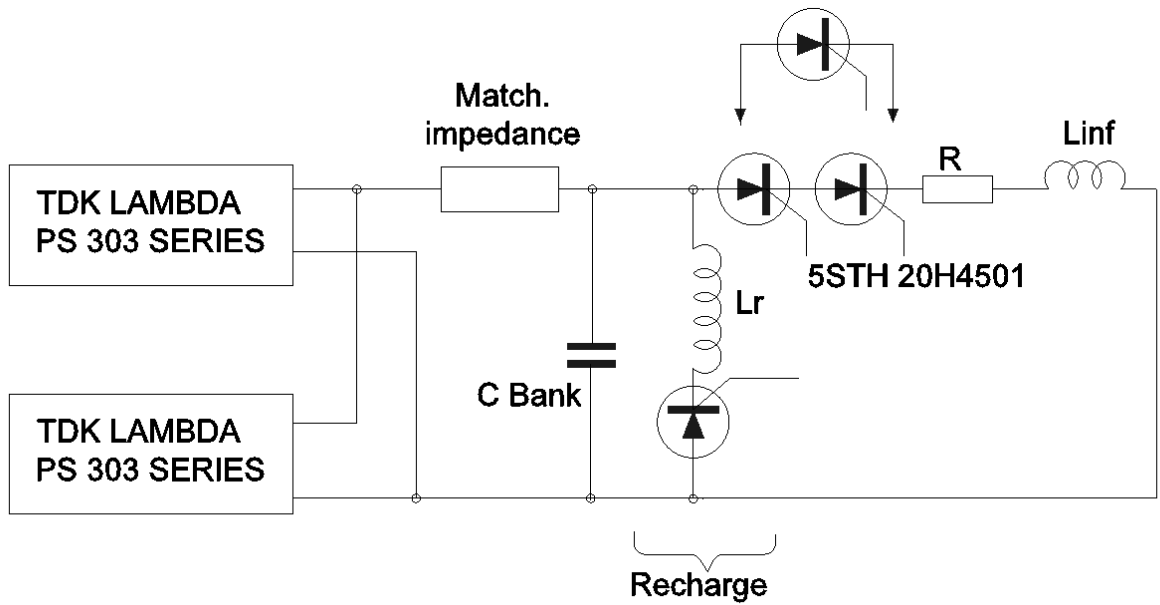

**Figure 3.** Electrical scheme. Resistor $R$ defines the discharge regime of capacitor.

One natural way to move discharge in oscillatory regime is in increasing the $R_w = \sqrt{L/C}$. If we fix the current value at *I=75kA*, then minimal voltage should be $V_{min} = IR_w = I\sqrt{L/C}$ @4.7 *kV*. With two thyristors in series the voltage could be ~ two times higher, up to 9 *kV*, so the $R_w$ could be ~doubled. This means that inductance could be twice bigger as before and capacitance could be two times smaller. Further optimization is possible.

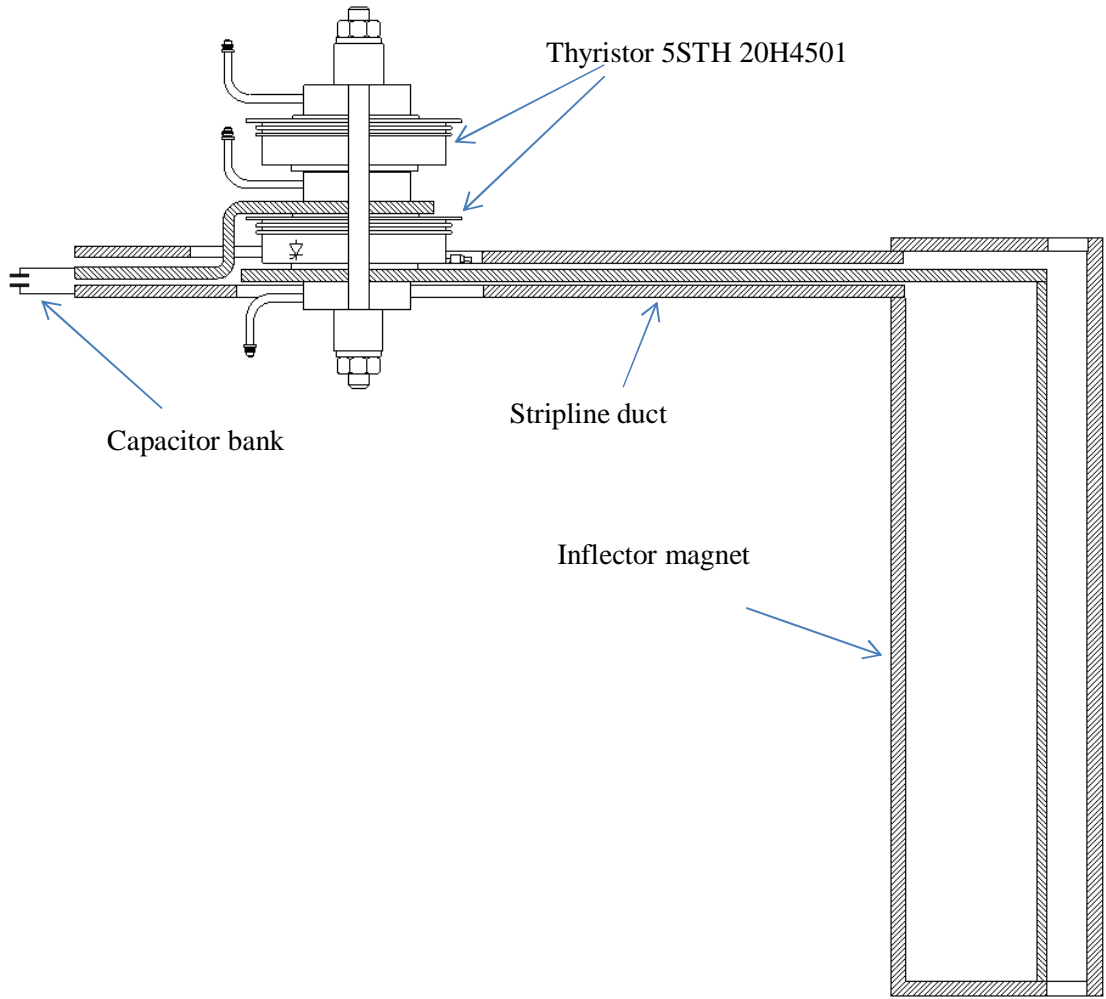

**Figure 4.** Installation of thyristor(s). Shown is the scheme with two thyristors in series interlaced by water-cooling tablets. Diameter of thyristor~90*mm*. Inflector length is shown reduced. Cooling channels for inflector conductors are not shown.

*Feeding with transformer* is another possibility. In Fig. 5. there is represented a transformer with low-stray inductance. It is wound with sections of coaxial cables; the outer braids of cables are connected in parallel and the inner conductor of primary winding runs over all sections. The transformer ratio 10:1 allows usage of thyratron as a switch. CX1725 (70*kV*, 15*kA*) could be used here.

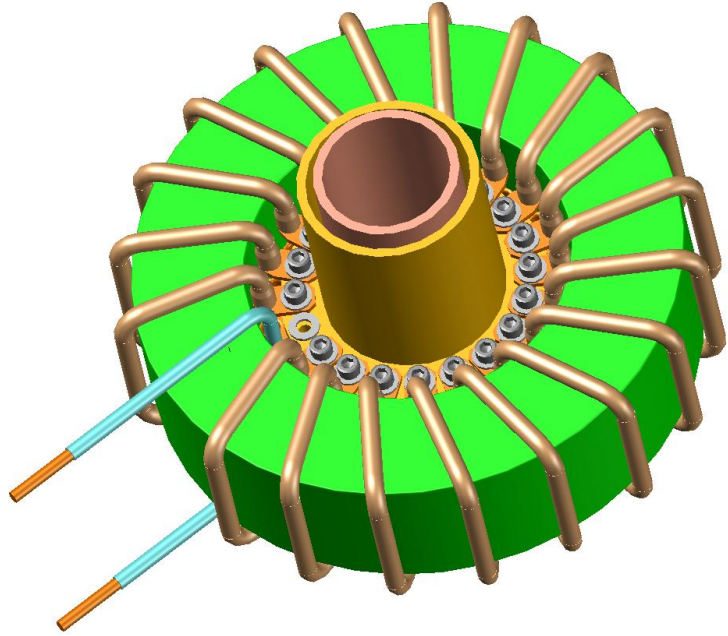

**Figure 5.** 3D view of low-inductive transformer.

One peculiarity here is that transformer could not transform a single-polarity pulse, so a bi-polar pulse appears in a secondary circuit which has period about two times shorter, than the primary half-sine pulse.

## DESIGN

This inflector magnet located in the same vacuum chamber, which used for SC inflector. Sketch of inflector magnet is represented in Fig.6. . Scheme with single-thyristor is shown in Fig.6 for simplicity. Middle part of inflector, feeding conductor and case could be made from solid block of hard Copper, so there will be no problems with attachment of central electrode to the case at the end. Feeding conductor (central in a stripline) supported by solid insulator insertions made from G11.

Central conductor and the adjacent wall could be made profiled for generation of gradient component of field, so the influence of passage of the muon beam through inhomogeneous field from inner side of magnet yoke towards central orbit could be reduced, see Fg.7. These gradient sections could be arranged at the end section of inflector only. The inclining magnetic field which bends the muon beam traversing through the channel towards the center of g-2 ring, has effective radial defocusing action (particles going at the left edge of transverse profile deflected lesser, than the ones which are going at the centroid of the beam), i.e. the fringe field is *focusing* vertically.

In Fig.7a), the central conductor is flat, so it practically not focusing beam in vertical direction (for central region, apart from vertical edge). For the profile in Fig.7 b) the inflector is focusing vertically, and for the profile Fig.7 c) it de-focuses the beam vertically, what is required for compensation of action of fringe field of main magnet.

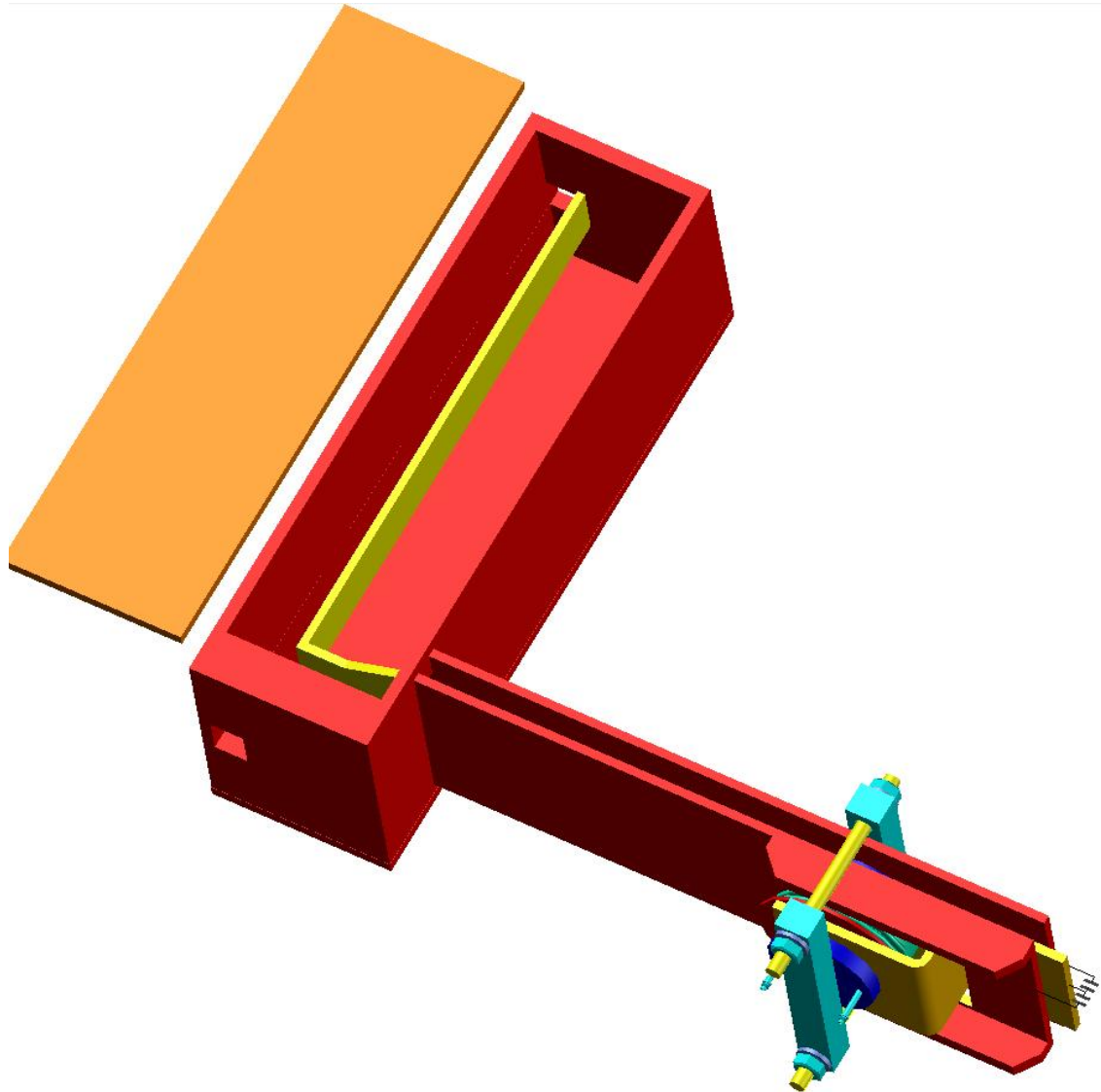

**Figure 6.** 3D sketch of Inflector magnet. Upper cover is lifted. Length of inflector and duct is shown reduced. Inner support fixtures are not shown.

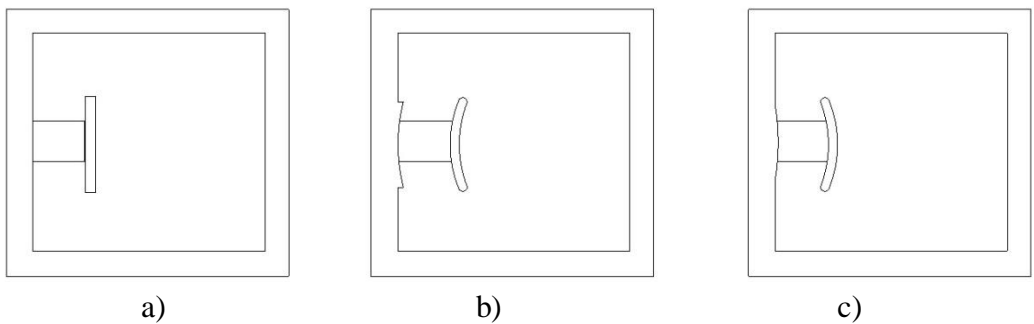

**Figure 7.** Possible transverse profiles of inflector. The center of g-2 ring located at the left. The inflector field itself deflects the beam to the right, as the main g-2 ring field deflects the beam to the left .

**10**

This magnet located inside Al vacuum chamber of g-2 ring in a place of SC inflector. Scheme with single-thyristor is shown in Fig.6 for simplicity. In principle the inflector magnet could be made slightly curved for better adjustment of the angle at the output of inflector orifice.

As the Copper case with pulsed magnet should fit into existing chamber, some adjustments should be done. First, as the same SC input duct will be used for allocation of the feeding stripline, the end region will be slightly different from that represented in Fig.1, more likely as it is shown in Fig.8.

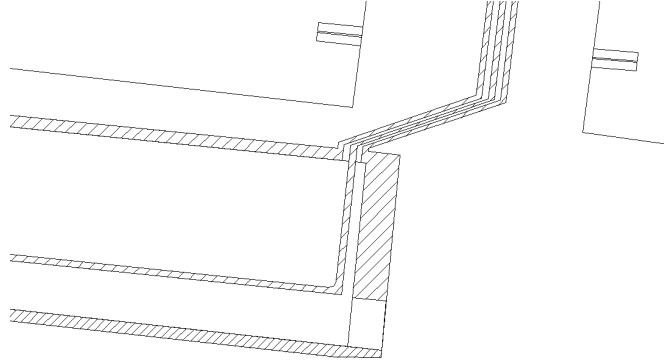

**Figure 8.** End region adjusted to fit the input to existing vacuum chamber.

## CONCLUSIONS

In this brief report we illuminated just a little amount of problems associated with pulsed inflector. We concluded, however, that the pulsed inflector is a feasible option and additional considerations are desirable. Thickness of septum in pulsed inflector (~10*mm*) is ~two times smaller than in SC one (23*mm*). Radial aperture could be increased up to 30 *mm* if necessary.

Reduction of aspect ratio *a/d* in Fig.2 by increasing the radial aperture and lowering the vertical one is extremely desirable option, as it reduces the current, required for inflector feeding. Probably inflector aperture *a*x*d*=3*cm*x3*cm* is the best one. As the septum in pulsed inflector has a thickness ≤1*cm*, while SC inflector's septum is 23*mm* thick, additional shift of injected beam centroid comes to ~5m*m* only. So the amplitude of kick should be bigger in 82/77=1.065 i.e.6.5%, what could be compensated by the kicker easily. So even lower aspect ratio *a/d* could be considered as well. We hope that to the time of possible implementation of pulsed inflector, the problems with the beam passage through first electrostatic quadrupole will be resolved positively, so there will be no restriction to enlargement of horizontal aperture of inflector. Inflector magnet could have a gradient and could be curved.

Once again, the basic concept of suggested inflector associated with usage of *thick-walls* surrounding the pulsed inflector magnet for protection of the field leakage into orbit by a skin-effect.

| | | | |
|---|---|---|---|
| $V_{DRM}$ | = | 4.5 kV | |
| $V_{RRM}$ | = | 18 V | |
| $I_{PULSE}$ | = | 80 kA | |
| $V_{Dcmax}$ | = | 2.8 kV | |

# High current high di/dt switch for Pulsed Power Applications
# 5STH 20H4501

## Features

- Asymmetric design
- For single or repetitve pulse applications
- Very high di/dt capability
- Free Floating Silicon Technology
- Glazed Ceramic Presspack Housing
- High interdigitated gate stucture
- Optimized as Closing Switch
- High Reliability

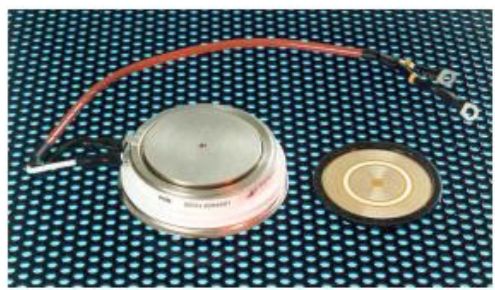

## Electrical Data

| | | | | |
|---|---|---|---|---|
| $V_{DRM}$ | Repetitive peak of-state voltage | 4.5 | kV | Tj = 125°C |
| $V_{RRM}$ | Repetitive reverse blocking voltage | 18 | V | Tj = 125°C |
| $V_{DC}$ | Permanent DC voltage for 100 FIT failure rate | 2.8 | kV | At $T_j \leq 125$ °C. Ambient cosmic radiation at sea level in open air. |
| $I_{PULSE}$ | Max. Pulse Current | 80 | kA | Half sine wave, Tj ≤ 50°C, tp ≤ 250 µs |
| di/dt | Max. current rate of rise | 18 | kA/µs | 1 Hz |
| $V_{GT}$ | Max. Gate trigger voltage | 4.0 | V | di$_G$/dt (min) 100 A/µs  Tj = 25° |
| $I_{GT}$ | Recomm. Gate trigger current | 120 | A | t = 20 µs |
| $V_t$ | Voltage drop | 2.36 | V | Tj = 50°C, I$_T$ = 3000 A |
| | | 2.42 | V | Tj = 125°C, I$_T$ = 3000 A |
| $I^2 t$ | Limiting load integral | 0.8 x10$^6$ | A$^2$s | tp = 250 µs, Tj = 50 °C |
| $V_{T0}$ | Threshold voltage | 1.28 | V | Tj = 50°C / Tj = 125°C |
| $r_T$ | Slope resistance | 0.36 | mΩ | Tj = 50°C |
| | | 0.38 | mΩ | Tj = 125°C |
| $t_{don}$ | Turn-on delay | ≤ 0.9 | µs | |

Same device without coaxial gate cable, but fast-on pins: P/N 5STH 2045H0001

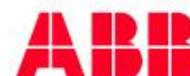

ABB Switzerland Ltd reserves the right to change specifications without notice